\documentclass[aps,pra,showpacs, longbibliography]{revtex4-2}

\usepackage{graphicx}
\usepackage{dcolumn}
\usepackage{bm}
\usepackage{color}
\usepackage{amsmath}
\usepackage{amssymb}
\usepackage{comment}
\usepackage{xcolor}
\newtheorem{theorem}{Theorem}
\begin{document}


\title{Demonstration of a bosonic quantum classifier with data re-uploading}

\author{Takafumi Ono$^{1,2}$}
 \email{ono.takafumi@kagawa-u.ac.jp}

\author{Wojciech Roga$^{3}$}
 \email{wojciech.roga@keio.jp}

\author{Kentaro Wakui$^{4}$}

\author{Mikio Fujiwara$^{5}$}

\author{Shigehito Miki$^{6}$}

\author{Hirotaka Terai$^{6}$}

\author{Masahiro Takeoka$^{3}$}
 \email{takeoka@elec.keio.ac.jp}

\affiliation{%
~\\
$^{1}$Program in Advanced Materials Science
Faculty of Engineering and Design,
Kagawa University,
2217-20 Hayashi-cho, Takamatsu, Kagawa
761-0396, Japan
}

\affiliation{%
$^{2}$JST, PRESTO, 4-1-8 Honcho, Kawaguchi, Saitama 332-0012, Japan
} 

\affiliation{%
$^{3}$Department of Electronics and Electrical Engineering, Keio University, 3-14-1 Hiyoshi, Kohoku-ku, Yokohama 223-8522, Japan\\
} 

\affiliation{%
$^{4}$National Institute of Information and Communications Technology (NICT), Koganei, Tokyo 184-8795, Japan
} 

\affiliation{%
$^{5}$Advanced ICT Research Institute, National Institute of Information and Communications Technology (NICT), Koganei, Tokyo 184-8795, Japan
}

\affiliation{%
$^{6}$Advanced ICT Research Institute, National Institute of Information and Communications Technology, 588-2 Iwaoka, Nishi, Kobe 651-2492, Japan
}

\date{\today}

\begin{abstract}
In a single qubit system, a universal quantum classifier can be realised using the data-reuploading technique. In this study, we propose a new quantum classifier applying this technique to bosonic systems and successfully demonstrated it using silicon optical integrated quantum circuits. We established a theory of quantum machine learning algorithm applicable to bosonic systems and implemented a programmable optical circuit combined with an interferometer. Learning and classification using part of the implemented optical quantum circuit with uncorrelated two-photons resulted in a classification with a reproduction rate of approximately 94\% in the proof of principle experiment. As this method can be applied to arbitrary two-mod N-photon system, further development of optical quantum classifiers, such as extensions to quantum entangled and multi-photon states, is expected in the future.

\end{abstract}

\maketitle
In recent years, there has been a lot of work on quantum machine learning algorithms that can apply machine learning methods to quantum systems in a sophisticated and efficient manner \cite{Paparo2014,Georgescu2014,Harris2017,Biamonte2017,Shen2017,Yu2019,Lloyd2020,Sweke2021}. Among quantum machine learning algorithms, classifiers belong to the supervised learning category and classify the input data into certain categories (classes) \cite{Mitarai2018,Schuld2020,Chen2021}. It is particularly effective for problems for which the correct or optimal answer is known. Usually, quantum classification consists of three steps: preparation of the quantum state, encoding of the data into the quantum state and measurement, whereby the parameters characterising the classifier are adjusted to build a classification model. Various quantum classifiers have been proposed, using quantum circuit models based on qubits \cite{Mitarai2018,Havlicek2019} and neural network models based on boson systems \cite{Lewenstein1991,Rebentrost2018,Steinbrecher2019,Cong2019}. In particular, the properties of these classifiers strongly depend on the data encoding method, and their performance depends on the number of qubits in the quantum circuits or the number of neurons in the hidden layer in the neural network, indicating that implementing higher-performance classifiers requires complex quantum circuits or networks \cite{Cerezo2021,Schuld2021}.

In this context, a universal quantum classifier using a single qubit has recently been developed by means of data re-uploading which allowed for nonlinear dependencies needed in a nontrivial learning process \cite{Perez-Salinas2020}. In this method, the unitary transform representing the quantum circuit is divided into multiple layers, and data-encoding and training are performed in each layer repeatedly. It has been shown that a universal quantum classifier can be realised with single qubit by increasing the number of layers of the circuit instead of increasing the number of qubits. The key idea of this method is to load the classical data sequentially into the unitary parameters representing the quantum circuit, rather than storing them distributed (or in parallel) to the quantum circuit \cite{Rebentrost2018,Cong2019,Steinbrecher2019}. As we will show later, this approach is applicable to qubit systems as well as bosonic systems even with a single photon which manifests remarkable computational power of photons working in an appropriate circuit. This motivates us to leverage the advantages of integrated photonics to develop a quantum classifier \cite{Carolan2015}.

Among the bosonic quantum circuit, silicon-based optical integrated circuits have the advantages of high integration density, sophisticated fabrication techniques and low optical losses in the communication wavelength band \cite{Thomson_2016}. In particular, it works well with CMOS electronics, allowing light to propagate on a waveguide while the quantum state of light is controlled by electrical circuits. There have been reports of on-chip quantum entangled state generation and measurement \cite{Raffaelli2016}, as well as the realisation of an optical interferometers operating with extinction ratios exceeding approximately 60 dB \cite{WILKES2016}, and the performance of each device has been improved \cite{Tasker2021}. The reason behind these developments of optical quantum integrated circuits is that it is feasible to construct programmable optical quantum circuits by combining multiple stages of interferometers \cite{Tillmann2013,Qiang2018,Paesani2019}. For these reasons, research into the application of programmable photonic quantum circuits has been intensified in recent years \cite{Wang2020}.

In this letter we report on the experimental realisation of a universal bosonic quantum classifier using silicon photonic circuits. We have applied the idea of a data re-uploading method proposed in qubit systems to bosonic quantum systems and implemented it using programmable integrated photonic circuits combined with multiple interferometers. As a proof-of-principle experiment, we used a two-mode circuit with a three-layer and two-photon input state as a bosonic system. For the optimisation of the circuit parameters, we used the sequential minimal optimisation method, which has been proposed in recent years \cite{Nakanishi2020}. Learning and classification using a part of the circuit successfully classified points divided by an elliptical boundary with a reproducibility of about 94\%. This work is expected to lead to further progress in optical quantum classifiers based on bosonic systems using quantum entangled and multi-photon states.

The universality argument for a single qubit processor provided by \cite{Perez-Salinas2020} relies on properties of SU2 group of unitary transformations that form the circuit. The group can be naturally parameterised by coefficients $\vec\phi$ in front of Pauli matrices $\vec \sigma$ which are the group generators, i.e., $U(\vec\phi)=e^{i\vec\phi\,\vec\sigma}$. Concatenation of two elements of SU2, $U(\vec\phi_1)$ and $U(\vec\phi_2)$ stays in the group, but its parameterisation is in general a non-linear function $\omega$ of the coefficients of the parts, $U(\vec\phi_1)U(\vec\phi_2)=U(\omega(\vec \phi_1,\vec \phi_2))$. Ref. \cite{Perez-Salinas2020} shows that using concatenation of many unitary transformations the parameters of which depend on the data $x$ as $\vec \phi_i=\vec \theta_i+\vec w_i x$, one can achieve the transformation with parameterization $U(\sum_i f(\vec \theta_i+ \vec w_i x))$, with a nonlinear function $f$. According to the universality theory for single layer neural networks the argument of this $U$ can approximate arbitrary function. This suggests that $U$ and functions of its elements can be complicated enough to realize desired computational tasks when sufficiently long circuit with data reuploeading is applied. In the two-mode bosonic circuit any SU2 transformation can be realized by passive elements and can define an input-output relation between modes creation operators. Therefore, the reasoning that leads to achievability of $U(\sum_i f(\vec \theta_i+ \vec w_i x))$ holds in this context as well. Also in this case the measurement leads to functions of the entries of the final $U$ therefore the computational power of the two mode bosonic processor is at least the same that of the one-qubit processor. We conjecture that two mode circuits with multiple photons may lead to some computational advantages.

\begin{figure}[t]
 \centering
 \includegraphics[keepaspectratio,scale=0.6]{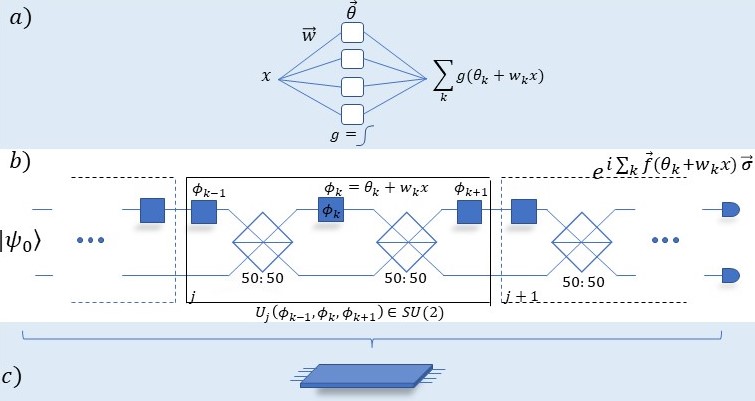}
 \caption{(a)A scheme of a single layer neural network charactrised by neurons with weights $\vec w$, biases $\vec \theta$ and a sigmoid activation function $g$. If the number of neurons is sufficiently large the output signal can approximate arbitrary function of data $x$. (b) Scheme of two-mode optical system realizing data reuploading scheme of a universal classifier. It consists of 50/50 beamsplitters and phase shifters characterised by phases $\phi_k$ which depend on data $x$ and adjustable parameters $w_k$ and $\theta_k$. One layer of the circuit encircled by the solid frame can realize arbitrary SU(2) transformation. (c) The optical system can be realized on silicon chip.
}
 \label{proposal}
\end{figure}

Here we adopted the data-reuploading method by \cite{Perez-Salinas2020} to bosonic system of the photonic quantum circuit. A schematic diagram is shown in Fig. \ref{proposal} where we indicate the analogy between two mode optical circuit and a classical single layer neural network that according to the universality theory is able to approximate an arbitrary function of the input data $x$ by its output signal $\sum_k g(\theta_k+w_kx)$, where $g$ is a nonlinear, usually sigmoid, function, $\vec w$ is a vector of weights and $\vec \theta$ is a vector of biases. The data re-uploading method decomposes the unitary operation of an optical quantum circuit into multiple sequential layers of unitary transformations the parameters of which depend on layer specific free parameters and the same data applied in each layer,
\begin{equation}
    \hat{U}(\vec{\phi}, \vec{x}^{(i)}) = \hat{U}_1(\vec{\phi_1}, \vec{x}^{(i)}) \hat{U}_2(\vec{\phi_2}, \vec{x}^{(i)})...\hat{U}_N(\vec{\phi_N}, \vec{x}^{(i)}).
\end{equation}
Here the index $i$ indicates one of data points. We define an arbitrary quantum state as $|\psi_0 \rangle $ which is input state of this quantum circuit. The input data $\{ \vec{x}^{(i)} \}$ to be classified is encoded into the parameter of the unitary operation of the circuit. Classification is then realized by optimizing the free parameters of each unitary operation of the circuit. The unitary operation for encoding and tuning is therefore a function of the input data $\{ \vec{x}^{(i)} \}$ and the tuning parameters $\vec{\phi}$ as given by $\hat{U}(\vec{\phi}, \vec{x}^{(i)})$. Finally, the probability $p^{(i)}(\vec\phi) = \left| \langle m|\hat{U}(\vec{\phi}, \vec{x}^{(i)}) |\psi_0 \rangle \right|^2$ of obtaining a particular output $m$ is measured. At this point, a threshold value of $b$ is set and the data is classified, for example, as yes if the probability $p^{(i)}(\vec\phi)$ is higher than the threshold value and no if it is lower. The tuning parameter $\vec{\phi}$ for training optical quantum circuits is optimised so that the least square error cost function $C(\vec{\phi})$ 
\begin{equation}
    C(\vec{\phi}) = \sum_i{\left( p^{(i)}(\vec{\phi}) - y^{(i)} \right)^2}\ 
\end{equation}
is as small as possible. Here $y^{(i)}$ is the classification of the training data, $y^{(i)}\in \{0,1\}$.

In the training process to adjust free parameters of the circuit we apply a modiﬁcation of gradient descent algorithm \cite{Sweke2020}, so-called Sequential Minimal Optimization (SMO) introduced in \cite{Nakanishi2020} adapted to our photonic set up 
\cite{Supplementary}. This approach relies on analytical minimization of the cost function, what under certain constraints, makes it more efficient than numerical methods such as the gradient method \cite{Nakanishi2020}.

\begin{figure*}[t]
 \centering
 \includegraphics[keepaspectratio,scale=0.5]{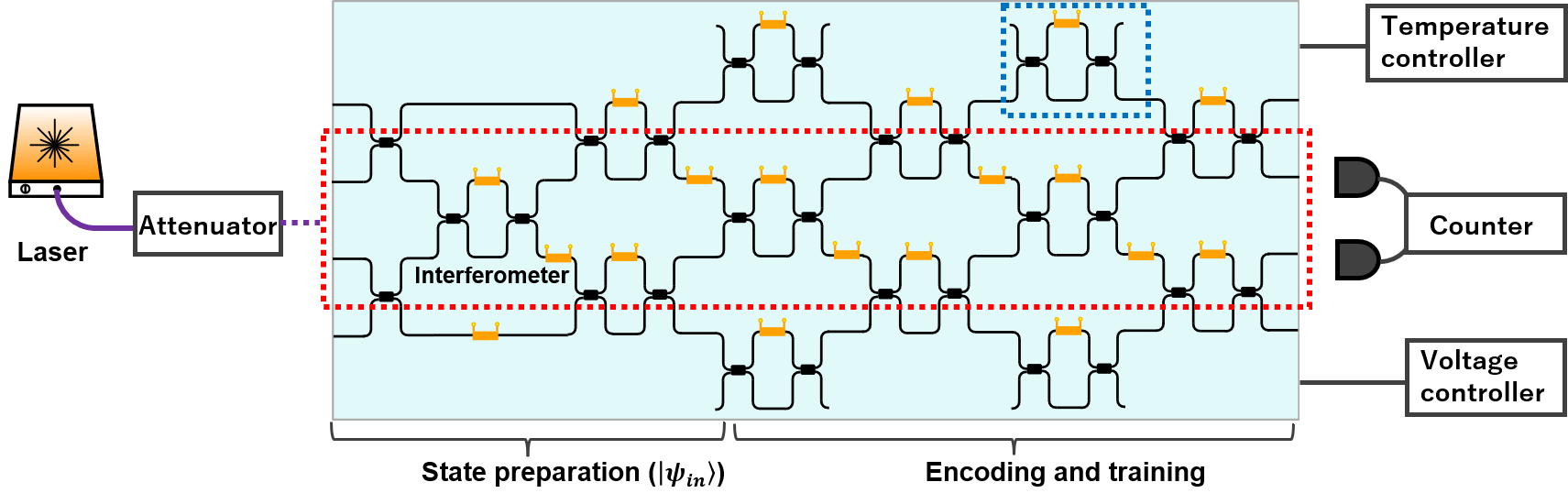}
 \caption{Schematic of the implemented optical quantum circuit. The first stage generates an arbitrary two-mode two-photon quantum state. The parameters for data and tuning are encoded into the parameters of the unitary transformation in the latter stage of the optical quantum circuit. The spot size converter for coupling the optical fibre to the optical waveguide is located at the top of the picture. The voltage to control the phase of the light is via electrical pads placed at the bottom of the picture. 
}
 \label{circuit}
\end{figure*}

The encoding of the input and the optimisation of the tuning parameters are performed repeatedly in each layer.
Note that in the present implementation, 
encoding of the input data into the quantum state and tuning of the circuit are performed simultaneously. 
In the present experiment, the optical quantum circuit was optimised so that the cost function is as small as possible by repeating this step several times, with the optimisation of the parameter $\vec{\phi}$ from layer $1$ to layer $N$ as a single step. In quantum circuit models, it has been shown that the value of the cost asymptotically approaches a minimum value when the number of steps is repeated, but the asymptotics of the minimum value in bosonic systems is a subject for future research.

We have implemented the quantum machine learning algorithm using silicon optical integrated quantum circuits. A schematic of the circuit is shown in fig. \ref{circuit}. The optical quantum circuits were fabricated using lithographic techniques on standard SOI wafers using a commercially available Multi Project Wafer service. The left-hand side is the input and the right-hand side is the output. The chip measures 2 mm x 3.5 mm and is placed on a copper substrate. The chip is temperature stabilised to operate at 25°C (stability 0.002 K) using a Peltier element. On one side of the chip, spot size converters are placed for input and output, and light from the optical fibre is coupled to the chip via the spot size converters. All of these fibres coupled to the chip without any problems and the coupling efficiency from fibre to chip was found to be approximately 4 dB. On the other side of the chip, electrical pads are placed, each with a voltage controller applying voltage via two adaptors (PCB boards).

First, a two-mode two-photon quantum state $|\psi_0 \rangle$ is generated using an interferometer and phase shifter. The quantum state is then manipulated by an arbitrary unitary transformation $\hat{U}(\vec{\phi}, \vec{x}^{(i)})$ using the middle part of the optical quantum circuit. The photons output from the chip are coupled via a spot size converter into an optical fibre and the photon count rate is detected using a superconducting photon detector. The quantum efficiency of superconducting photon detectors is around 70\% to 80\%. 

The middle stage optical quantum circuit consists of four inputs and four outputs and combines an interferometer and a phase shifter to realise a programmable optical quantum circuit. The interferometer consists of two beam splitters and one phase shifter. The beam splitters are implemented using spatial multimode interference by widening the silicon optical waveguide until the propagating light becomes spatially multimode. Control of the phase of the light was achieved by applying voltage to a thermal resistive element placed above the optical waveguide and locally applying heat to change the refractive index of the silicon optical waveguide. Note that some irreducible interferometers have been added to ensure proper alignment of the phase shifter inside the circuit (e.g. the interferometer surrounded by the blue line in the circuit.) The interferometers were evaluated using laser light and the extinction ratio was measured to be more than 30 dB for all interferometers. There were a total of 27 phase shifters placed on the chip and all were confirmed to work properly. Experimental data on the dependence of the phase shifter on the applied voltage is shown in fig. \ref{circuit}(c). When the voltage applied to the thermal resistors was varied from 3 V to 8 V, the phase of the light changed in the range of 0 to 4 $\pi$.


\begin{figure}[t]
 \centering
 \includegraphics[keepaspectratio,scale=0.4]{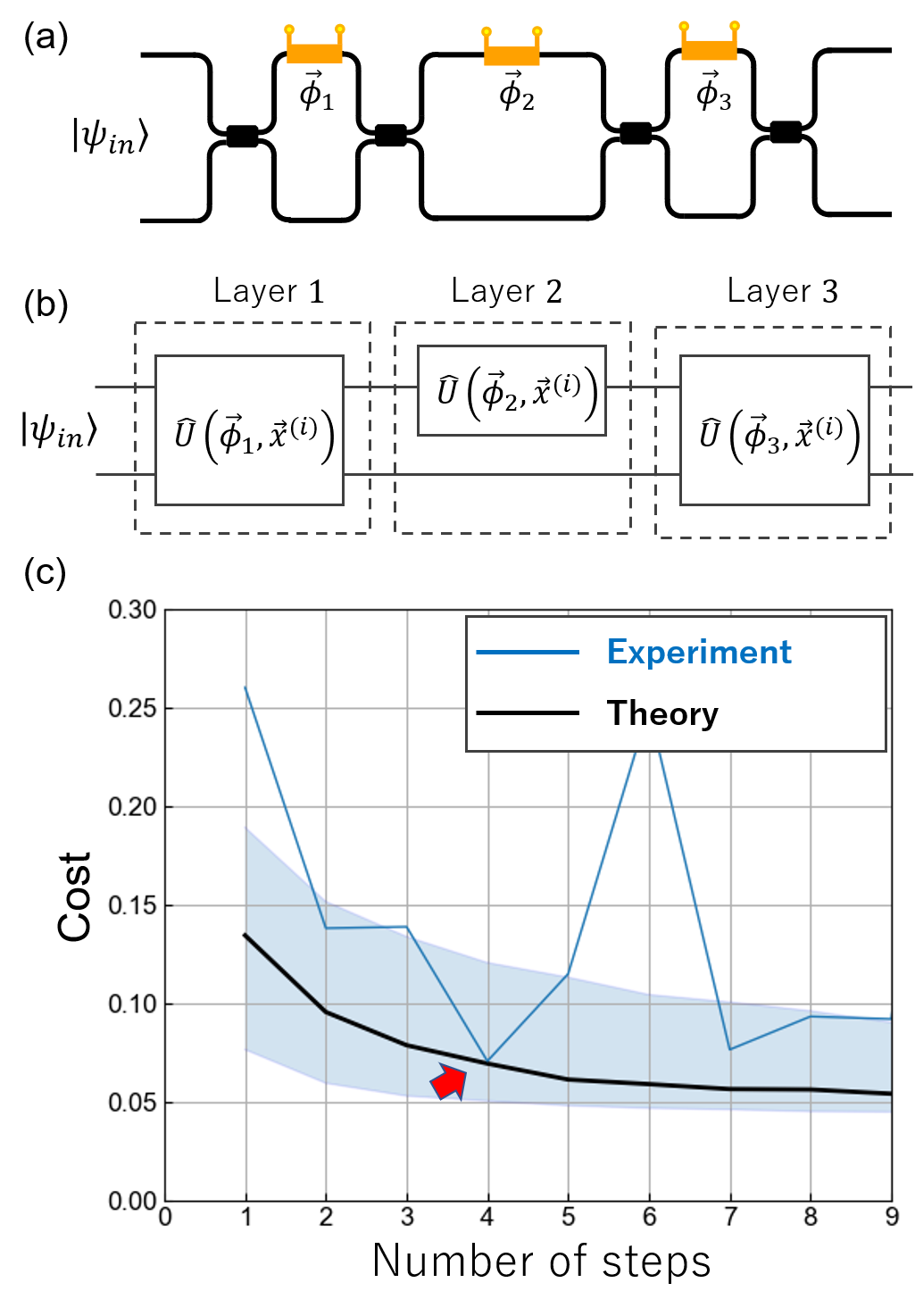}
 \caption{(a) Schematic of the photon circuit used. The circuit shown in the red box in Fig. \ref{circuit}(a) is decomposed into three layers. (b) Value of the cost at each step, optimised using the data reuploading technique. The black line is the theoretical value and the blue line is the experimental value. The theoretical values are the average of the values optimised by simulation using 200 training data (3000 times in total). The region between the 10th and 90th percentile quartiles is shown in light blue.
}
 \label{simplified_circuit}
\end{figure}
In this experiment, we used a part of the circuit (enclosed by the red dotted line) for an optical quantum classifier, and also used uncorrelated two-mode two-photon state as an input. Note, however, that the method proposed in this study can be used for any quantum state. A simplified quantum circuit is shown in Fig. \ref{simplified_circuit}(a). We prepared arbitrary uncorrelated two-mode two-photon state as $|\psi_{in} \rangle_{ab}$, where $a$ and $b$ are the spatial modes of the photons. The quantum circuit consists of three layers; $\hat{U}_1(\varphi_1)$ and $\hat{U}_3(\varphi_3)$ are the unitary operations of a Mach-Zehnder interferometer, which are the functions of $\varphi_1$ and $\varphi_3$, respectively, and the $\hat{U}_2(\varphi_2)$ is a unitary operation that gives the optical phase shift between two spatial modes of $a$ and $b$. 
We used the two dimensional data as $\vec{x} = (x_1,x_2)$ to be classified. 
Data $\vec{x}$ and tuning parameter $\vec{\phi}$ are encoded into the parameters of unitaries as follows,
\begin{eqnarray}
\varphi_1 &=& \phi_1 + \phi_2 \times x_2 \nonumber\\
\varphi_2 &=& \phi_3 + \phi_4 \times x_1 \nonumber\\
\varphi_3 &=& \phi_5 + \phi_6 \times x_2, \nonumber
\end{eqnarray}
and so on.
The output quantum state in the version of the experiment with three layers after the whole circuit is therefore given by
\begin{equation}
| \psi_{out} \rangle = \hat{U}_3(\varphi_3) \hat{U}_2(\varphi_2) \hat{U}_1(\varphi_1) |\psi_{0} \rangle.
\end{equation}
and can be easily extended in more complicated versions.
The measurements were performed with one photon output for each spatial mode for each photon ($\langle m|_{ab} = \langle 1;1|_{ab}$). As a result, the probability obtained at the output of the optical quantum circuit is given by
\begin{equation}
    p(\vec{\phi}, \vec{x}) = \left| \langle 1;1 | \hat{U}_3(\varphi_3) \hat{U}_2(\varphi_2) \hat{U}_1(\varphi_1) |\psi \rangle_{0} \right|^2.
\end{equation}
For a given 2D function, the tuning parameters $\vec{\phi} = (\phi_1, \phi_2, \phi_3, \phi_4, \phi_5, \phi_6)$ are optimised so that the cost given by Eq.(1) becomes as small as possible. In the present study, we find an explicit formula for $p(\vec{\phi}, \vec{x})$ for each $\phi$ in each step of the training by using the SMO method \cite{Supplementary} instead of minimising the cost numerically by gradient descent. The method we applied is described in the supplementary material \cite{Supplementary}.

\begin{figure}[t]
 \centering
 \includegraphics[keepaspectratio,scale=0.35]{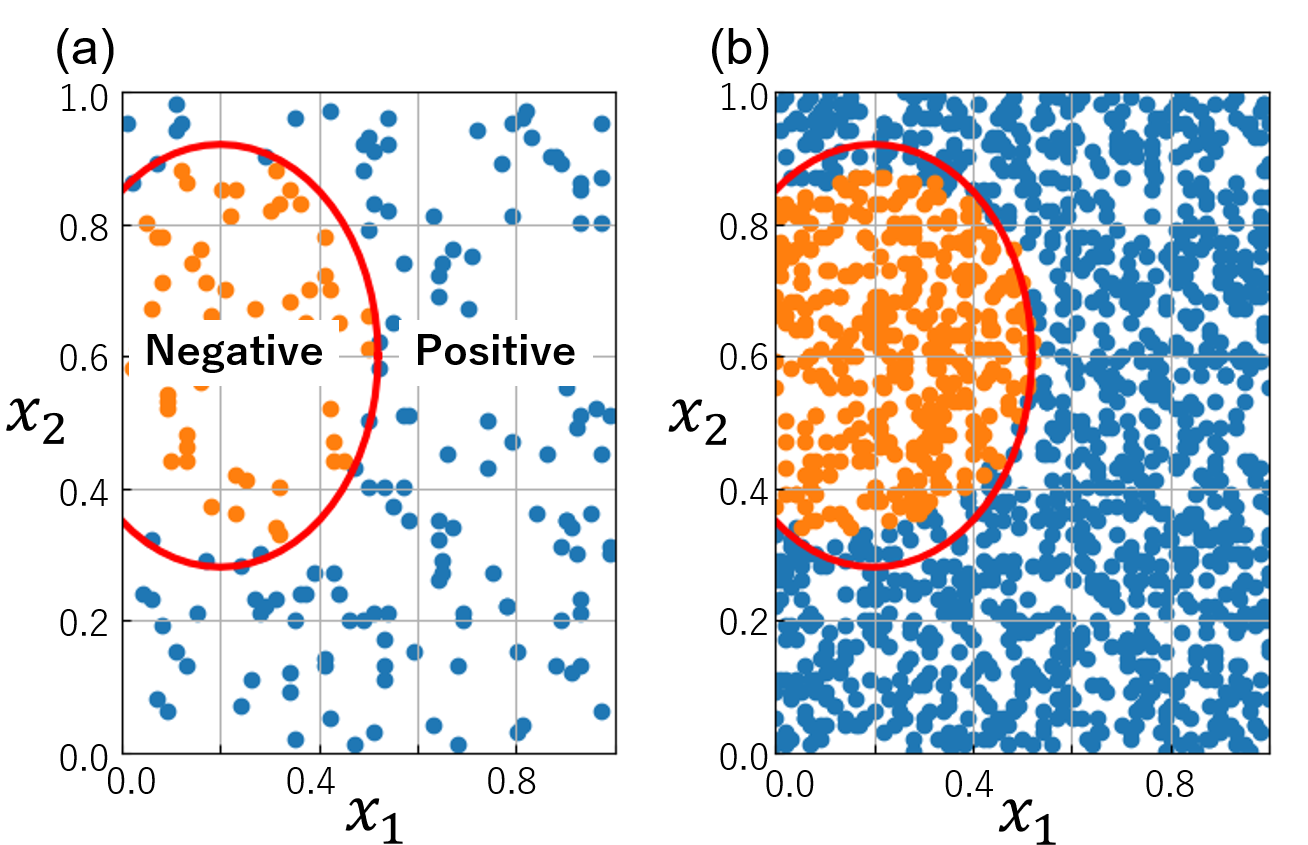}
 \caption{(a) The area to be classified in the light quantum circuit is shown. The red lines are the boundaries of the regions to be divided. (b) Classification results. Yellow dots are teacher data belonging to negative regions, blue dots are teacher data belonging to positive regions.
}
 \label{result}
\end{figure}

In this experiment, we chose the circle specified by $x_1$ and $x_2$ as a function for classification,
\begin{equation}
    (x_1 - 0.2)^2 + (x_2 - 0.6)^2 = 0.32^2.
\end{equation}
The classification was then carried out with training data points inside of this circle classified as $y^{(i)}=0$ and points outside of this circle classified as $y^{(i)}=1$. The number of training data points used was 200, and the parameter $\vec{\phi}$ was optimised to keep the cost as low as possible using the SMO method. Figure \ref{simplified_circuit} (b) shows the optimal value of the cost at each step. The black lines are the results from theoretical simulations; the blue lines are experimental values. Theoretical simulations confirm that the cost decreases monotonically with each step, as expected. On the other hand, in the experiments, there was a tendency for costs to increase at certain steps, so the classification was carried out under the condition that costs were lowest at all steps in the experiments. The reason why we observe the jump is the numerical instability of the algorithm used to find the minimum of the cost function.

Figure \ref{result} shows the results of classification on 1500 inputs using the trained optical quantum circuit. It was confirmed that the classification was visually correct. Table 1 shows the results of our classification. We obtain true positive(TP) and false negative(TN) as 1085 and 10, respectively. We also obtain false positive(FP) and true negative(TN) as 45 and 360, respectively. We then calculated true positive rate(TPR) and true negative rate(TNR) as 
\begin{eqnarray}
    TPR &=& \frac{TP}{TP + FN} = 99.1 \% \nonumber\\
    TNR &=& \frac{TN}{TN + FP} = 88.9 \%.
\end{eqnarray}
Thus, on average, the classification was confirmed to be correct with a correct response rate of approximately 94\%.

In conclusion, we have applied the universal quantum classifier method proposed for qubit systems to optical quantum circuits in boson systems. Specifically, the unitary transformation of the optical quantum circuit was decomposed into multiple layers using the Data reuploading technique. In each layer, the data and tuning parameters were encoded into the parameters of the unitary transform. Machine learning techniques were then used to optimise costs. We implemented the proposed optical quantum classifier by mounting an optical quantum circuit with a total of about 220 optical components on a silicon optical waveguide. The quantum states prepared are uncorrelated two-mode two-photon states and the number of classes is 2. 200 training data were used to train the optical quantum circuit and performing classification on 1500 test data, the quantum classifier was implemented with a correct response rate of approximately 94\%. Although the quantum states used in this study are not entangled, research on quantum classifiers using quantum entangled states is a subject for future work possible on the same circuit. In general, we have shown that the quantum classifier with data reuploading method proposed for cubit systems can be naturally applied to boson-based optical quantum circuits. In the future, it is expected that increasing the number of spatial modes and the number of photons used will lead to higher-performance quantum classifiers.

\textbf{\emph{Acknowledgements}}
This work was supported by PRESTO, Japan Science and Technology Agency and The Murata Science Foundation.

\bibliography{reference}

\newpage

\section{Appendix: Sequential minimum optimization}

We consider a two-mode two-photon circuit with a few layers of unitary transformations. Some of them perform data re-uploading analogous to \cite{Perez-Salinas2020}. Each layer is a unitary transformation that can be decomposed into two single phase shifters and two 50-50 beamsplitters \cite{Clements2016}. 

So, without loss of generality we can assume that the parameters of the circuit are phases of the phase shifters. Without loss of generality we assume also that at the end we measure the probability of the state $|11\rangle$ from which we infer the class of a data-point. The cost function will be a function of this probability. Let us write the probability of measuring $|11\rangle$ distinguishing one of the variables $\phi$ to be adjusted during training,
\begin{equation}
p(\phi) = |\langle 11|U_2^\dagger U_\phi U_1|\Psi_0\rangle|^2 .
\label{prob}
\end{equation}
Here, $|\Psi_0\rangle$ is a fixed initial state, $U_1$ and $U_2$ are unitaries before and after a phase shifter $U_\phi$ parameterised by $\phi$. The idea is to update each parameter $\phi$ to the value where the cost function expressed as the function of this parameter achieves its minimum. This technique known as sequential minimum optimization was introduced before in \cite{Nakanishi2020}, however we introduce here our our notation and methodology. 
The sequential minimum optimization requires an explicit formula for $p(\phi)$ for each $\phi$ in each step of the training. To do that let us expand $U_1|\Psi_0\rangle$ and $U_2|11\rangle$ in the Fock basis
\begin{eqnarray}
U_1|\Psi_0\rangle &=& \alpha_1|20\rangle + \beta_1|11\rangle + \gamma_1|02\rangle,\\
U_2|11\rangle &=& \alpha_2|20\rangle + \beta_2|11\rangle + \gamma_2|02\rangle.
\end{eqnarray}
In this basis 
\begin{equation}
    U_\phi = \begin{pmatrix}e^{2i\phi}&0&0\\0&e^{i\phi}&0\\0&0&1\end{pmatrix}.
\end{equation}
Hence, from (\ref{prob}) it follows easily that
\begin{equation}
p = A_0 + A_1\cos(\phi) + A_2 \sin(\phi) + A_3 \cos(2\phi) + A_4 \sin(2\phi),
\label{proba}
\end{equation}
where
\begin{eqnarray}
A_0 &=& |\alpha_1|^2|\alpha_2|^2+|\beta_1|^2|\beta_2|^2+|\gamma_1|^2|\gamma_2|^2,\label{Azero}\\
A_1 &=& 2\Im(\bar\alpha_1\alpha_2\beta_1\bar\beta_2) + 2\Im(\bar\beta_1\beta_2\gamma_1\bar\gamma_2),\\
A_2 &=&-2\Re(\bar\alpha_1\alpha_2\beta_1\bar\beta_2)-2\Re(\bar\beta_1\beta_2\gamma_1\bar\gamma_2),\\
A_3 &=&2\Im(\bar\alpha_1\alpha_2\gamma_1\bar\gamma_2),\\ 
A_4 &=&2\Re(\bar\alpha_1\alpha_2\gamma_1\bar\gamma_2).\label{A4}
\end{eqnarray}
If we know $A_0,...,A_4$ we have $p(\phi)$. Values $A_i$ can be obtained by solving the set of 5 equations given by (\ref{proba}) measured for 5 different $\phi$
\begin{equation}
\begin{pmatrix}
1&\cos\phi_0&\sin\phi_0&\cos2\phi_0&\sin2\phi_0\\
1&\cos\phi_1&\sin\phi_1&\cos2\phi_1&\sin2\phi_1\\
1&\cos\phi_2&\sin\phi_2&\cos2\phi_2&\sin2\phi_2\\
1&\cos\phi_3&\sin\phi_3&\cos2\phi_3&\sin2\phi_3\\
1&\cos\phi_4&\sin\phi_4&\cos2\phi_4&\sin2\phi_4
\end{pmatrix}  
\begin{pmatrix}
A_0\\A_1\\A_2\\A_3\\A_4
\end{pmatrix}=
\begin{pmatrix}
r_0\\r_1\\r_2\\r_3\\r_4
\end{pmatrix}.
\label{cosmatrix}
\end{equation}
Values $r_k$ are the probabilities measured by the same circuit in chosen phases of the investigated element. Choosing proper values of $\phi_k$ this formula reduces to a Fourier transform. Namely, if 
\begin{eqnarray}
\phi_0 &=& 0,\\
\phi_1 &=& \frac{2\pi}{5},\\
\phi_2 &=& -\frac{2\pi}{5},\\
\phi_3 &=& \frac{4\pi}{5},\\
\phi_4 &=& -\frac{4\pi}{5}
\end{eqnarray}
equation (\ref{cosmatrix}) becomes 
\begin{equation}
F  
\begin{pmatrix}
(A_3 - iA_4)/2\\(A_1-iA_2)/2\\A_0\\(A_1 + iA_2)/2\\(A_3+iA_4)/2
\end{pmatrix}=
\begin{pmatrix}
r_3\\r_1\\r_0\\r_2\\r_4
\end{pmatrix},
\label{omegamatrix}
\end{equation}
where the Fourier transform matrix is
\begin{equation}
   F = \begin{pmatrix}
\omega^4&\omega^2&\omega^0&\omega^{-2}&\omega^{-4}\\
\omega^2&\omega^1&\omega^0&\omega^{-1}&\omega^{-2}\\
\omega^0&\omega^0&\omega^0&\omega^0&\omega^0\\
\omega^{-2}&\omega^{-1}&\omega^0&\omega^{1}&\omega^{2}\\
\omega^{-4}&\omega^{-2}&\omega^0&\omega^{2}&\omega^{4}\\
\end{pmatrix},
\label{fourier}
\end{equation}
and
\begin{equation}
    \omega = e^{i\frac{2\pi}{5}}.
\end{equation}
The inverse of the Fourier matrix is 
\begin{equation} 
F^{-1} = \frac{1}{5}F^{\dagger}.
\end{equation}
Therefore,
\begin{equation}
\begin{pmatrix}
A_3\\A_1\\A_0\\A_2\\A_4
\end{pmatrix}= \frac{1}{5}SF^\dagger 
\begin{pmatrix}
r_3\\r_1\\r_0\\r_2\\r_4
\end{pmatrix},
\end{equation}
where matrix $S$ is a simple change of variables,
\begin{equation}
S = \begin{pmatrix}
1&0&0&0&1\\
0&1&0&1&0\\
0&0&1&0&0\\
0&i&0&-i&0\\
i&0&0&0&-i\\
\end{pmatrix}.
\end{equation}
Explicitely,
\begin{equation}
\begin{pmatrix}
A_3\\A_1\\A_0\\A_2\\A_4
\end{pmatrix}= \frac{2}{5}
\begin{pmatrix}
\cos\frac{8\pi}{5}&\cos\frac{4\pi}{5}&1&\cos\frac{4\pi}{5}&\cos\frac{8\pi}{5}\\
\cos\frac{4\pi}{5}&\cos\frac{2\pi}{5}&1&\cos\frac{2\pi}{5}&\cos\frac{4\pi}{5}\\
0.5&0.5&0.5&0.5&0.5\\
\sin\frac{4\pi}{5}&\sin\frac{2\pi}{5}&0&-\sin\frac{2\pi}{5}&-\sin\frac{4\pi}{5}\\
\sin\frac{8\pi}{5}&\sin\frac{4\pi}{5}&0&-\sin\frac{4\pi}{5}&-\sin\frac{8\pi}{5}
\end{pmatrix}
\begin{pmatrix}
r_3\\r_1\\r_0\\r_2\\r_4
\end{pmatrix}.
\end{equation}

This shows that to get the functional dependence of the probability $p(\phi)$ in (\ref{proba}) on each of the adjustable circuit parameters $\phi$, we need to measure 5 probabilities $r_k$ with the same circuit but with different values of the parameter under investigation. 

Notice that with the effort of finding $A_k$ for each training point we can find explicit formula for the cost function and its minimum for each adjustable parameter. Hence, we can update the parameter with the value in which the cost function achieves the minimum. 

In our application, for each training point numerated by index $i$, for each phase shifter characterised by its $\phi$, we measure five probabilities for five phases as discussed above. From the probabilities we calculate parameters $A_k^i$ that allow us to define function $p^i(\phi)$ as in (\ref{proba}) that enters to the cost function
\begin{equation}
    C(\phi) = \frac{1}{N}\sum_{i=1}^N (p^i(\phi) - y^i)^2.
\end{equation}
This is still an easy function of $\phi$ that can be easily minimized numerically. We upgrade the phase of the phase shifter as follows
\begin{equation}
    \phi\rightarrow {\rm argmin}\ C(\phi),
\end{equation}
and we move to the next phase shifter. We repeat everything until a stopping criterion is met.

\end{document}